\documentclass[12pt]{article}
\usepackage{graphicx}
\usepackage{epsfig}
\usepackage{epstopdf}
\DeclareGraphicsExtensions{.pdf,.eps,.png,.jpg,.mps}

\def\lsim{\mathrel{\rlap{\lower3pt\hbox{\hskip0pt$\sim$}}
     \raise1pt\hbox{$<$}}}         
\def\gsim{\mathrel{\rlap{\lower4pt\hbox{\hskip1pt$\sim$}}
     \raise1pt\hbox{$>$}}}         

\usepackage{amsmath}
\usepackage{amsfonts}

\begin{document}
\begin{titlepage}

\centerline{\Large \bf Coping with Negative Short-Rates}
\medskip

\centerline{Zura Kakushadze$^\S$$^\dag$\footnote{\, Zura Kakushadze, Ph.D., is the President of Quantigic$^\circledR$ Solutions LLC,
and a Full Professor at Free University of Tbilisi. Email: \tt zura@quantigic.com}}
\bigskip

\centerline{\em $^\S$ Quantigic$^\circledR$ Solutions LLC}
\centerline{\em 1127 High Ridge Road \#135, Stamford, CT 06905\,\,\footnote{\, DISCLAIMER: This address is used by the corresponding author for no
purpose other than to indicate his professional affiliation as is customary in
publications. In particular, the contents of this paper
are not intended as an investment, legal, tax or any other such advice,
and in no way represent views of Quantigic® Solutions LLC,
the website \underline{www.quantigic.com} or any of their other affiliates.
}}
\centerline{\em $^\dag$ Free University of Tbilisi, Business School \& School of Physics}
\centerline{\em 240, David Agmashenebeli Alley, Tbilisi, 0159, Georgia}
\medskip
\centerline{(February 9, 2015; revised: August 7, 2015)}

\bigskip
\medskip

\begin{abstract}
{}We discuss a simple extension of the Ho and Lee model with generic time-dependent drift in which: 1) we compute bond prices analytically; 2) the yield curve is sensible and the asymptotic yield is positive; and 3) our analytical solution provides a clean and simple way of separating volatility from the drift in the short-rate process. Our extension amounts to introducing one or two reflecting barriers for the underlying Brownian motion (as opposed to the short-rate), which allows to have more realistic time-dependent drift (as opposed to constant drift). In our model the spectrum -- or, roughly, the set of short-rate values contributing to bond and other claim prices -- is discrete and positive. We discuss how to calibrate our model using empirical yield data by fitting three parameters and then read off the time-dependent drift.
\end{abstract}
\medskip
\end{titlepage}

\newpage

\section{Introduction}

{}An allure of short-rate models is their apparent elegance and simplicity. In theory, once the short-rate process $r_t$ and the corresponding risk neutral measure are specified, bonds and other claims, such as bond options, can be priced via simple-looking conditional expectations of ``exponentially" discounted claims. However, in practice, things are a bit trickier. In their simplest incarnations, {\em e.g.}, Merton's (1973) model and the Ho and Lee (1986) model, short-rate models typically\footnote{\, One can circumvent this via apparently unrealistic time dependence for the drift in $r_t$.} produce unrealistic yield curves for zero-coupon bonds, with long-maturity yields turning negative.

{}An immediately evident -- but not necessarily relevant -- culprit would appear to be that in such models $r_t$ can become negative. However, intuitively, since $r_t$ is not observable in real life, it taking negative values need not be a big deal. Instead, what is important is that the model yield curve be sensible. Thus, in the Vasicek (1977) model $r_t$ can become negative, but for a range of parameters the model yield curve {\em a priori} appears sensible as the spectrum in this model is discrete and bounded from below. In models such as Merton's model and the Ho and Lee model the spectrum is continuous and unbounded. Intuitively, the ``spectrum" here can be thought of as the set of values of $r_t$ that actually contribute to the bond (and other claim) prices.\footnote{\, Mathematically, the ``spectrum" is the set of eigenvalues for the static Schr\"odinger equation to which the pricing PDE reduces, {\em e.g.}, when the underlying parameters are time-independent.}

{}While the spectrum in the Vasicek model is bounded from below, it is nonnegative only if, roughly speaking, volatility is low. In some cases, including in the current low interest rate environment with not-so-low volatility, this can be problematic. In fact, one can ensure that $r_t$ is always positive and can never reach zero as, {\em e.g.}, in the Black and Karasinski (1991) model, where $\ln(r_t)$ is the Vasicek process. However, here too, low interest rates by construction imply low (almost vanishing) volatility.

{}An alternative approach to ensure $r_t\geq 0$ is to introduce a reflecting barrier at $r_t=0$, or to treat $r_t$ as an option on an underlying ``shadow" rate $X_t$ (which can be negative) by taking its positive part,\footnote{\, For a reflecting barrier at $r_t = 0$ we have a Neumann boundary condition w.r.t. $r_t$ ($r_t\in {\bf R}^+$) and the system can be described via $r_t = |X_t|$ with claims symmetric under $X_t\rightarrow -X_t$. The spectrum is discrete and positive even if $X_t\in {\bf R}$ follows Merton's model or the Ho and Lee model (see below). For $r_t$ as an option we have $r_t=(X_t)^+$ (with the ``shadow" rate $X_t\in {\bf R}$), and at $r_t=0$ we sew together solutions with $r_t =0$ ($X_t\in {\bf R}^-$) and $r_t = X_t \in {\bf R}^+$ by requiring continuity of the claim price and its first derivative w.r.t. $X_t$. Here the $r_t = X_t\in {\bf R}$ model is assumed to have a discrete spectrum, and then the spectrum in the $r_t = (X_t)^+$ model is discrete and positive.} as in (Black, 1995) and (Rogers, 1995, 1996). Short-rate models with a reflecting barrier at $r_t=0$ were studied in (Goldstein and Keirstead, 1997) for Merton's model, the Vasicek model, the extended (a.k.a. shifted) CIR (Cox, Ingersoll and Ross, 1985) model, and the Longstaff (1989) model. Models with $r_t$ as an option were studied in (Gorovoi and Linetsky, 2004) when the ``shadow" rate $X_t$ follows the Vasicek model and the shifted CIR model. In this approach volatility generally need not be small even in the low interest rate regime.

{}In all the models studied in (Goldstein and Keirstead, 1997) and (Gorovoi and Linetsky, 2004) the underlying parameters (volatility, drift, mean-reversion rate, {\em etc.}) are assumed to be constant. More generally, in time-homogeneous cases (where the parameters depend on $r_t$ (or $X_t$) but have no explicit time dependence), we have the standard separation of variables and the pricing PDE reduces to an ODE (static Schr\"odinger equation). Also, at $r_t = 0$ we have a time-independent boundary condition for a reflective barrier and a time-independent sewing condition for $r_t$ as an option. These nice simplifying features are lost once we consider time-dependent parameters, which in many cases are required to describe real-life yield curves.

{}However, not all is lost. In this paper we set forth a simple way of circumventing this difficulty. The following observations will pave the way for us. First, to get a sensible yield curve (and, more generally, claim pricing), it is not required that $r_t$ be nonnegative.\footnote{\, Thus, the models studied in (Goldstein and Keirstead, 1997) and (Gorovoi and Linetsky, 2004) can have: a reflecting barrier at $r_t = r_- < 0$; and $r_t = r_- + (X_t)^+$ with $r_- < 0$, respectively.} What is required is that the spectrum be discrete and nonnegative.\footnote{\, Realistically, the lowest eigenvalue has to be reasonably positive (see below).} To achieve this, it suffices to introduce a reflecting barrier for the underlying Brownian motion $W_t$ (as opposed to $r_t$). Then we can have time-dependent drift and still achieve separation of variables. Furthermore, the resulting boundary condition is also time-independent. Second, with a reflecting barrier for $W_t$, the spectrum is discrete and positive (roughly, if $r_t>r_-<0$)\footnote{\, {\em I.e.}, for this purpose alone we do not need more complex (Vasicek, CIR, {\em etc.}) dynamics.} already in the Ho and Lee model\footnote{\, More generally, we discuss factorized processes of the form $r_t = U(W_t) + \chi(t)$. Furthermore, if desired, we can have $r_t\geq 0$ by appropriately choosing parameters. However, this is not required.}
\begin{equation}
 dr_t = \sigma~dW_t + \nu(t)~dt
\end{equation}
where we take constant volatility $\sigma$ and general time-dependent drift $\nu(t)$ and solve the pricing problem analytically.\footnote{\, Both for one reflecting barrier (say, at $W_t = 0$) and two reflecting barriers (at $W_t = 0$ and $W_t = L$). The latter case can be useful for preventing $r_t$ from wandering away into large values.} Because of the simplicity of the diffusion part in the Ho and Lee model, our analytical solution provides a clean and simple way of separating volatility in $r_t$ from the drift. We discuss how to fit this model into an empirical yield curve, which involves calibrating three constant parameters (volatility $\sigma$, the initial value of $r_t$, and the location of the reflecting barrier) and then read off the time-dependent drift, or, equivalently, its contribution to the yield curve. We illustrate our method by fitting the model into recent U.S. Treasury yield data.

{}The remainder of this paper is organized as follows. In Section 2 we briefly review some generalities of short-rate models and then discuss how to introduce reflecting barriers for short-rate processes of the form $r_t = U(W_t) + \chi(t)$ and solve the claim pricing problem. In Section 3 we apply the results of Section 2 to the Ho and Lee model with one and two reflecting barriers, give explicit formulas for zero-coupon bond prices, and for illustrative and comparison purposes discuss fitting the model into the Japanese Government Bond data used to calibrate a model in (Gorovoi and Linetsky, 2004). We also discuss a fit into recent U.S. Treasury yield data. We briefly conclude in Section 4. Some technical details are relegated to Appendices.

\section{Short-rate Models}

{}A short-rate model posits a risk-neutral measure {\bf Q} and a short-rate process $r_t$. The cash bond process is given by
\begin{equation}
 B_t=\exp\left(\int_0^t r_s~ds\right)
\end{equation}
while the price at time $t$ of a claim $X$ at maturity $T$ is given by $v(r_t, t, T)$, where the pricing function $v(z, t, T)$ is given by
\begin{equation}\label{v_t}
 v(z,t,T)\equiv
 \left\langle\exp\left(-\int_t^T r_s~ds\right) X\right\rangle_{{\bf Q},r_t=z}
\end{equation}
{\em E.g.}, for $X=1$, we have $v(r_t,t,T)=P(t,T)$, and $v(z,T,T)=1$, where $P(t,T)$ is the price of a zero-coupon $T$-bond. Also, $\langle\cdot\rangle$ denotes expectation.

{}In short-rate models one usually works with a parameterized family of processes, and chooses the parameters to best fit the market.
Thus, typically one {\em assumes} that $r_t$ satisfies the following SDE:
\begin{equation}\label{r.markov}
 dr_t=\sigma(r_t,t) dW_t+\nu(r_t,t) dt
\end{equation}
where $\sigma(z,t)$ and $\nu(z,t)$ are deterministic functions, and $W_t$ is a ${\bf Q}$-Brownian motion. The pricing function $v(z, t, T)$ satisfies a pricing PDE\footnote{\, Known as the Feynman-Kac equation, a.k.a. the Arrow-Debreu security PDE.}, which follows from the requirement that the discounted process
$Z(t,T)\equiv B_t^{-1}v(r_t,t,T)$ be a martingale under the risk-neutral measure ${\bf Q}$:
\begin{equation}\label{pricing.PDE}
 \nu(r_t,t) \partial_z v(r_t,t,T)+ \partial_t
 v(r_t,t,T)+{1\over 2}\sigma^2(r_t,t) \partial^2_z v(r_t,t,T)-r_t v(r_t,t, T)=0
\end{equation}
with the terminal condition $v(z,T,T)=\langle X\rangle_{{\bf Q},r_T = z} \equiv {\widetilde Y}(z)$.

\subsection{Mean-Reversion and Positivity}

{}In some cases one may wish to require that the short-rate process $r_t$ not wander away to large values. One way to achieve this is to use a mean-reverting process
\begin{equation}\label{OU}
 dX_t = \sigma(t) dW_t + \left[\theta(t) - \alpha(t) r_t\right] dt
\end{equation}
where $\sigma(t)$, $\theta(t)$ and $\alpha(t)$ depend only on time. For constant $\sigma$, $\theta$ and $\alpha$ we have the mean-reverting
Ornstein-Uhlenbeck process. In the Vasicek/Hull-White model we have $r_t = X_t$. One ``shortcoming" of this model is that $r_t$ can occasionally become negative.\footnote{\, More relevantly (see below), {\em e.g.}, for constant $\sigma$, $\theta$ and $\alpha$, the asymptotic ({\em i.e.}, large $T$) zero-coupon bond yield in this model is positive only for $\sigma^2 <  2\alpha\theta$.} One way to deal with this is to consider short-rate models of the form $r_t = r_0~f(X_t)/f(0)$, where $f(y)$ is a positive function,
{\em e.g.}, $f(y) = \exp(y)$, which is the Black-Karasinski model. The path integral treatment of such models was discussed in (Kakushadze, 2014).\footnote{\, The results we obtain below can also be derived using path integral.}

\subsection{Short-rate Models with Reflecting Barriers}

{}An alternative approach is to consider short-rate processes of the form $r_t = U(W_t, t)$, where the function $U(x, t)$ is bounded. More precisely, here we can consider three cases: i) $x$ is unrestricted, $x\in {\bf R}$; ii) $x$ takes values on a half-infinite line, $x\in [x_-,\infty)$ or $x\in(-\infty, x_+$]; and iii) $x$ takes values on a finite interval, $x\in [x_-,x_+]$. In the case iii), $U(x, t)$ can be a simple function, {\em e.g.}, $U(x,t) = a(t) x + b(t)$, even though it is not bounded when extended to the entire real line $x \in {\bf R}$. In the case ii), $r_t$ is nonnegative\footnote{\, Albeit, as discussed above and below, nonnegativity of $r_t$ is actually no longer required in cases ii) and iii) and is replaced by a weaker condition.} so long as $U(x,t)\geq 0$ for, say, $x\in [x_-,\infty)$.

{}In the following we will assume that, for the allowed values of $x$, for a given value of $t$, there is a one-to-one mapping between $z$ and $x$ via $z = U(x, t)$. Then the pricing PDE (\ref{pricing.PDE}) simplifies as follows. Let ${\widetilde Y}(U(x, T)) \equiv Y(x)$ and\footnote{\, When $U(x, t)$ has no explicit time dependence, we have time translational invariance and $\psi(x, t)$ depends on $T$ only via the combination $T-t$. However, this is no longer the case when $\partial_t U(x, t) \neq 0$. Nonetheless, below we will use the abbreviated notation $\psi(x, t)$.} $v(U(x, t), t, T) \equiv \psi(x, t)$. Also, we have $\sigma(r_t, t) = \partial_x U(x, t)$ and $\nu(r_t, t) = \partial^2_x U(x, t)/2 + \partial_t U(x, t)$, where $U(x, t) = r_t$, and the pricing PDE (\ref{pricing.PDE}) reads:
\begin{eqnarray}\label{Sch}
 &&\partial_t \psi(x,t) = -{1\over 2}~\partial^2_x \psi(x,t) + U(x,t)~\psi(x,t)\\
 &&\psi(x,T) = Y(x)\label{boundary}
\end{eqnarray}
where (\ref{boundary}) is the terminal condition at $t=T$. For definiteness, here we focus on the case with two boundaries. Cases with a single (lower or upper) boundary follow upon removing the unwanted boundary to the corresponding infinity and requiring that $\psi(x,t)$ be finite (actually, vanish) at such infinity.

{}Since we have boundaries, we need to specify boundary conditions. For the reasons which will become evident momentarily, we will take the boundaries $x_\pm$ to be reflecting barriers, {\em i.e.}, when the Brownian motion touches the lower (upper) boundary $x_-$ ($x_+$) from above (below), it is reflected back into the values $x>x_-$ ($x<x_+$). This implies that we have Neumann boundary conditions
\begin{eqnarray}
 &&\partial_x \psi(x_-,t) = 0\\
 &&\partial_x \psi(x_+,t) = 0
\end{eqnarray}
This then implies that $Y(x)$ also satisfies Neumann boundary conditions:
\begin{eqnarray}
 &&\partial_x Y(x_-) = 0\\
 &&\partial_x Y(x_+) = 0
\end{eqnarray}
Had we imposed the Dirichlet boundary conditions $\psi(x_-,t)=0$ and $\psi(x_+,t)=0$, for which the process $W_t$ is not allowed to touch the boundaries, we would invariably have $Y(x_-)=0$ and $Y(x_+)=0$. This would not be suitable for our purposes here as Dirichlet boundary conditions are incompatible with, {\em e.g.}, the claim $X=1$ for a zero-coupon $T$-bond.\footnote{\, We could consider inhomogeneous boundary conditions such that $Y(x)$ would not need to satisfy Neumann boundary conditions. However, this will not be needed for our purposes here.}

{}Note that (\ref{Sch}) is the Schr\"odinger equation in imaginary (Euclidean) time for a particle (with mass $m=1$ and Planck's constant $\hbar=1$) in the potential $U(x, t)$. For general time-dependent potentials $U(x, t)$ Eq. (\ref{Sch}) is difficult to solve. However, for our purposes here it will suffice to consider factorized potentials of the form
\begin{equation}\label{factor}
 U(x, t) = V(x) + \chi(t)
\end{equation}
For such potentials we have $\sigma(r_t, t) = V^\prime(x)$ and $\nu(r_t, t) = V^{\prime\prime}(x)/2 + \dot{\chi}(t)$, where a prime denotes a derivative w.r.t. $x$, while a dot stands for a derivative w.r.t. $t$, and $r_t = V(x) + \chi(t)$.

{}For the factorized potential (\ref{factor}), we have separation of variables, and the solution to (\ref{Sch}) can be written in terms of a series ($E_1 < E_2 < E_3 <\dots$):\footnote{\, Here we are assuming that the eigenvalue spectrum $E_n$ is discrete and bounded from below. This is the case if the potential $V(x)$ is confining, even without the boundaries. With one or two boundaries, it suffices that $V(x)$ is bounded from below for the allowed range of $x$.}
\begin{eqnarray}
 &&\psi(x, t) = e^{-\eta(t,T)}\sum_{n=1}^\infty c_n~\psi_n(x)~e^{-\chi_n(t) \left(T-t\right)}\\
 &&\eta(t,T)\equiv \int_t^T ds \left[\chi(s) - \chi(t)\right]\label{eta}\\
 &&\chi_n(t)\equiv \chi(t) + E_n
\end{eqnarray}
where $\psi_n(x)$, $n\in {\bf N}$ is the complete orthonormal set of solutions to the static Schr\"odinger equation:
\begin{eqnarray}\label{Sch.stat}
 &&-{1\over 2}~\partial^2_x \psi_n(x) + V(x)~\psi_n(x) = E_n~\psi_n(x)\\
 &&\int_{x_-}^{x_+} dx~\psi_n(x)~\psi_{n^\prime}(x) = \delta_{nn^\prime}\label{norm}
\end{eqnarray}
subject to the Neumann boundary conditions
\begin{eqnarray}
 &&\partial_x \psi_n(x_-) = 0\\
 &&\partial_x \psi_n(x_+) = 0
\end{eqnarray}
If we have a single lower (upper) reflecting boundary at $x_-$ ($x_+$), then we have only one Neumann boundary condition at this boundary together with the requirement that $\psi_n(x)$ vanish as $x \rightarrow +\infty$ $(-\infty)$.

{}For a zero-coupon $T$-bond we have $X=1$ and $v(r_t, t, T) = P(t, T)$, where $P(t, T)$ is the bond price. The yield is given by
\begin{equation}
 R(t, T) = - {\ln(P(t,T))\over {T-t}}
\end{equation}
The asymptotic yield must be nonnegative\footnote{\, Realistically, the asymptotic yield should not be less than some positive number.}
\begin{equation}\label{asymp.yield}
 0 \leq R_*(t) \equiv \lim_{T-t\rightarrow +\infty} R(t,T) =  \chi_*(t) + \chi_1(t) = \chi_*(t) + \chi(t) + E_1
\end{equation}
where
\begin{equation}
 \chi_* \equiv \lim_{T-t\rightarrow +\infty} {\eta(t,T)\over {T-t}}
\end{equation}
Assuming $\chi_*(t)$ is finite, we have a requirement that $E_1 \geq -\chi_*(t) - \chi(t)$.

{}The coefficients $c_n$ (for a general claim $Y(x)$) are given by
\begin{equation}\label{c_n}
 c_n = \int_{x_-}^{x_+} dx~\psi_n(x)~Y(x)
\end{equation}
which is a consequence of (\ref{boundary}) and (\ref{norm}). To recap, the pricing function $v(z, t, T)$ is given by
\begin{equation}
 v(z, t, T) = e^{-\eta(t,T)} \sum_{n=1}^\infty c_n~\psi_n(x)~e^{-\chi_n(t)\left(T-t\right)}
\end{equation}
where $x$ is related to $z$ via $V(x) +\chi(t) = z$.

\section{Ho and Lee Model with Reflecting Barriers}

{}Let us consider the simplest example, the Ho and Lee model
\begin{equation}\label{HL}
 dr_t = \sigma~dW_t + \nu(t)~dt
\end{equation}
where $\sigma$ is constant, but the drift $\nu(t)$ {\em a priori} is an arbitrary function of $t$ (subject to some restrictions we discuss below). We have
\begin{equation}
 r_t = \sigma~W_t + \chi(t)
\end{equation}
where (we are assuming $W_0 = 0$)
\begin{equation}
 \chi(t) \equiv r_0 + \int_0^t ds~\nu(s)
\end{equation}
So, $V(x) = \sigma~x$.

{}The static Schr\"odinger equation (\ref{Sch.stat}) reads:
\begin{equation}\label{Airy}
 -{1\over 2}~\partial^2_x \psi_n(x) + \sigma~x~\psi_n(x) = E_n ~\psi_n(x)
\end{equation}
The solution depends on the boundary conditions and is expressed via a linear combination of $Ai\left(\alpha x - e_n\right)$ and $Bi\left(\alpha x - e_n\right)$, where $\alpha\equiv (2\sigma)^{1/3}$, $e_n\equiv E_n/\beta$, $\beta \equiv \sigma/\alpha = (\sigma^2/2)^{1/3}$, and the Airy functions $Ai(y)$ and $Bi(y)$ are the two independent solutions of the Airy equation $f^{\prime\prime}(y) = y~f(y)$.

\subsection{Ho and Lee Model on a Semi-infinite Line}

{}Thus, let us consider the Ho and Lee model (\ref{HL}) on a semi-infinite line, {\em i.e.}, we restrict the values of $W_t$ to ${\bf R}^+$, where without loss of generality we have set $x_-=0$. We have (\ref{Airy}) with only one Neumann boundary condition
\begin{equation}\label{N1}
 \partial_x\psi_n(0)=0
\end{equation}
and the requirement that $\psi_n(x)$ be finite as $x\rightarrow+\infty$. The solution is given by:
\begin{equation}\label{semi1}
 \psi_n(x) = a_n~Ai\left(\alpha~x-e_n\right)
\end{equation}
where
\begin{equation}
 e_n = -\xi_n
\end{equation}
Here $\xi_n$ ($0>\xi_1>\xi_2>\dots$) are the zeros of the first derivative of $Ai(y)$:
\begin{equation}
 Ai^\prime\left(\xi_n\right)=0
\end{equation}
Note that the integration in (\ref{c_n}) is from 0 to $+\infty$. The normalization coefficients
\begin{equation}
 a^{-2}_n \equiv \int_0^\infty dx~\left[Ai\left(\alpha x-e_n\right)\right]^2 = -\alpha^{-1}{\xi_n\left[Ai\left(\xi_n\right)\right]^2}\label{semi2}
\end{equation}
where we have used (\ref{int1}) (see Appendix \ref{appA}).

\subsubsection{``Modulus" Ho and Lee Model}

{}In the Ho and Lee model on a semi-infinite line the short-rate process $r_t$ generally is unbounded from above. If this is not problematic,\footnote{\, {\em E.g.}, for given volatility $\sigma$, we are interested in time horizons such that $r_t$ simply does not have enough time to wander away too far.} then we do not even need to restrict the process $r_t$ to a semi-infinite line. Instead, we can simply consider a ``modulus" model:\footnote{\, Note the difference with, {\em e.g.}, an alternative model $r_t = \left|\sigma W_t + \chi(t)\right|$. The latter is harder to tackle for general $\chi(t)$. See Appendix \ref{appB} for details.}
\begin{equation}\label{mod}
 r_t = \sigma~|W_t| + \chi(t)
\end{equation}
where $W_t$ is unrestricted. In this case we have
\begin{eqnarray}\label{mod1}
 &&\psi_n(x) = a_n~Ai\left(\alpha~x-e_n\right),~~~x\geq 0\\
 &&\psi_n(x) = (-1)^{n+1}\psi_n(-x),~~~x <0\\
 && a^{-2}_n \equiv 2\int_0^\infty dx~\left[Ai\left(\alpha~x-e_n\right)\right]^2\label{mod2}\\
 && a^{-2}_n = 2~\alpha^{-1}e_n\left[Ai\left(-e_n\right)\right]^2,~~~n=1,3,\dots\\
 && a^{-2}_n = 2~\alpha^{-1}\left[Ai^\prime\left(-e_n\right)\right]^2,~~~n=2,4,\dots
\end{eqnarray}
and $e_n$ are now given by
\begin{eqnarray}
 &&e_n = -\xi_{(n+1)/ 2},~~~n=1,3,\dots\\
 &&e_n = -\zeta_{n/2},~~~n=2,4,\dots
\end{eqnarray}
Here $\zeta_n$ ($0>\zeta_1>\zeta_2>\dots$) are the zeros of $Ai(y)$:
\begin{equation}
 Ai(\zeta_n) = 0
\end{equation}
Note that the integration in (\ref{c_n}) is now from $-\infty$ to $+\infty$.

{}The solutions $\psi_n(x)$ with odd $n$ are symmetric w.r.t. the ${\bf Z}_2$ reflections $x\rightarrow -x$ and satisfy the Neumann boundary condition (\ref{N1}), while the solutions with even $n$ are antisymmetric and satisfy the Dirichlet boundary condition $\psi_n(0)=0$. The short-rate process is invariant under $W_t \rightarrow -W_t$, so the claims should also be symmetric: $Y(-x)=Y(x)$. Then we have $c_n = 0$ for even $n$, and for odd $n$ we have
\begin{equation}
 c_n = 2\int_0^\infty dx~\psi_n(x)~Y(x)
\end{equation}
It then follows (taking into account the differing by $\sqrt{2}$ normalizations for $\psi_n(x)$ in the two cases) that, for the same (symmetric) claim $X=Y(W_T)$, the pricing function $v(z,t,T)$ is the same in the Ho and Lee Model on a semi-infinite line and in the ``modulus" model (\ref{mod}) -- as they should be based on symmetry considerations.

\subsection{Ho and Lee Model on an Interval}

{}Assuming the reflecting barriers at $x=0$ and $x=L$, the Brownian motion $W_t$ wanders between 0 and $L$. We have (\ref{Airy}) with two Neumann boundary conditions
\begin{equation}\label{N}
 \partial_x\psi_n(0) = \partial_x\psi_n(L) = 0
\end{equation}
The solution to (\ref{Airy}) is given by
\begin{eqnarray}\label{box1}
 &&\psi_n(x) = a_n~\phi_n(x)\\
 &&\phi_n(x)\equiv Ai\left(\alpha~x-e_n\right) - Q(0, e_n)~Bi\left(\alpha~x-e_n\right)\label{box5}\\
 &&Q(x, e) \equiv {Ai^\prime\left(\alpha~x-e\right)\over Bi^\prime\left(\alpha~x-e\right)}\label{box3}
\end{eqnarray}
where $e_n$ are the roots of the following equation for $e$:
\begin{equation}\label{box4}
 Q(L, e) = Q(0, e)
\end{equation}
which has an infinite tower of discrete solutions $e_n > 0$, $n\in {\bf N}$. The normalization coefficients
\begin{equation}\label{box2}
 a^{-2}_n \equiv \int_0^L dx~\phi_n^2(x) = {1\over\alpha}\left[e_n~\phi_n^2(0) + \left(\alpha~L - e_n\right)\phi_n^2(L)\right]
\end{equation}
where we have used (\ref{box4}) and (\ref{int1}) (see Appendix \ref{appA}). Note that if we take $L\rightarrow\infty$, (\ref{box2}) reduces to (\ref{semi2}), as it should.

\subsubsection{``Periodic" Ho and Lee Model}

{}Just as in the case of the Ho and Lee model on a semi-infinite line, the Ho and Lee model on an interval is also equivalent to a model where $W_t$ is unrestricted. In this unbounded model the short-rate process has the following ``periodic" form
\begin{equation}
 r_t = \sigma~h(W_t) + \chi(t)
\end{equation}
where $h(x)$ is a piecewise linear periodic function
\begin{eqnarray}
 &&h(x) = x,~~~0\leq x \leq L\\
 &&h(x) = 2L-x,~~~L\leq x \leq 2L\\
 &&h(x + 2L) = h(x)
\end{eqnarray}
and the claim $X = Y(W_T)$ is symmetric under both $x\rightarrow -x$ and $L+x\rightarrow L-x$ reflections: $Y(-x) = Y(x)$ and $Y(L-x) = Y(L+x)$. This implies that $Y(x)$ is periodic: $Y(x+2L) = Y(x)$. Therefore, the integration in (\ref{c_n}) is from $0$ to $2L$, or, equivalently, from $-L$ to $L$.

\subsection{Bond Pricing on a Semi-infinite Line}

{}For illustrative purposes, let us discuss zero-coupon $T$-Bond pricing, for which the claim is simply $X=1$. For the Ho and Lee model on a semi-infinite line we have
\begin{equation}\label{c1}
 c_n = \int_0^\infty dx~\psi_n(x) = {1\over\alpha}~a_n \int_{\xi_n}^\infty dy~Ai\left(y\right) = {\pi\over\alpha}~a_n~Ai(\xi_n)~Gi^\prime(\xi_n)
\end{equation}
where we have used (\ref{int2}) (see Appendix \ref{appA}), and $Gi(y)$ is a Scorer function.

{}The $T$-bond pricing function therefore is given by the following simple formula:
\begin{equation}\label{T-bond}
 v(z, t, T) = e^{-\eta(t, T)}\sum_{n=1}^\infty {\pi~Gi^\prime(\xi_n)\over |\xi_n|~Ai(\xi_n)}~Ai\left({z - \chi_n(t)\over \beta}\right) e^{-\chi_n(t)\left(T-t\right)}
\end{equation}
where
\begin{equation}
 \chi_n(t) = \chi(t) + E_n = \chi(t) + \beta\left|\xi_n\right|
\end{equation}
and $\eta(t, T)$ is defined in (\ref{eta}). Recall that $\beta = (\sigma^2/2)^{1/3}$.

{}Taking into account the asymptotics given in Appendix \ref{appA}, we see that the series is well-behaved at large $n$. Thus, let
\begin{equation}
 v(z, t, T) \equiv \sum_{n=1}^\infty v_n(z, t, T)
\end{equation}
For large $n$ such that $|\xi_n| \gg \left(z - \chi(t)\right)/\beta$, the leading asymptotics are:
\begin{eqnarray}
 &&v_n(z, t, T) \sim u_n~\exp\left(-\gamma\left(t,T\right)n^{2/3}\right)\\
 &&u_n \equiv (-1)^{n+1}~\sqrt{2\over 3n}\\
 &&\gamma(t,T)\equiv \beta\left(T-t\right)\left({3\pi\over 2}\right)^{2/3}
\end{eqnarray}
So, asymptotically, we have an alternating series, which converges according to the Leibniz criterion.\footnote{\, Note that $\sum_{n=1}^\infty u_n = \sqrt{2/3}~(1-\sqrt{2})~\zeta(1/2)\approx 0.494$, where $\zeta(s)$ is the Riemann zeta function.} In numerical computations one would truncate the series at a suitably chosen finite $n$ (see below).

\subsubsection{Parameter Count and Model Calibration}

{}Suppose we have data for zero-coupon\footnote{\, In practice, unless the yield data is readily available, we may have data for coupon-bearing bond prices, which then are bootstrapped to obtain zero-coupon bond prices.} $T$-bond prices. How many parameters would we need to fit to calibrate (\ref{T-bond})? We need to break this down. Let us start with the case of vanishing drift $\nu(t)\equiv 0$. Then the answer is that we have 3 parameters to fit. Indeed, in (\ref{T-bond}) we have $z$, $\beta$ (which is fixed by $\sigma$) and $r_0$ (note that $\chi(t)\equiv r_0$ when $\nu(t)\equiv 0$). However, looking at (\ref{pricing.PDE}) for this model with $\nu(t)\equiv 0$
\begin{eqnarray}\label{FK}
 &&\partial_t v(z,t,T) + {1\over 2}~\sigma^2~\partial_z^2 v(z,t,T) - z~v(z, t, T) = 0\\
 &&v(z,T,T) = 1\label{FK.T}
\end{eqnarray}
it might appear that we have only 2 parameters, $z$ and $\sigma$, to fit. The third parameter, $r_0$, is simply the location of the reflecting boundary,\footnote{\, When $\nu(t)\equiv 0$, the reflecting boundary at $x_-=0$ for the Brownian motion process $W_t$ translates into a reflecting boundary at $r_- = r_0$ for the short-rate process $r_t$. As discussed in more detail in Appendix \ref{appB}, this is no longer the case for a general time-dependent drift $\nu(t)$.} which is a free\footnote{\, Modulo the requirement (\ref{asymp.yield}), that is (see below). Note that $\chi_*(t) \equiv 0$ when $\nu(t)\equiv 0$.} parameter. In fact, in the Ho and Lee model, before introducing a reflecting boundary, the spectrum of (\ref{FK}) is continuous.\footnote{\, Moreover, (\ref{FK}) has a symmetry under the transformation $v(z,t,T) \rightarrow e^{-\zeta t}~v(z+\zeta, t, T) \equiv {\widetilde v}_\zeta(z, t, T)$,
where $\zeta$ is an arbitrary constant. {\em I.e.}, if $v(z,t,T)$ satisfies (\ref{FK}) and (\ref{FK.T}), then so does ${\widetilde v}_\zeta(z, t, T)$ for arbitrary $\zeta$.} Once we introduce a reflecting boundary, the spectrum is discrete and, once again, its nonnegativity does not require that the reflecting boundary be at $r_-=0$. It suffices to require that $r_-=r_0\geq -\beta\left|\xi_1\right|$.

{}When the drift $\nu(t)$ is nonzero, we have a choice. Thus, we can try to fit more parameters. {\em E.g.}, we can assume that $\nu(t) \equiv \nu_0\neq 0$ is constant; then we have four parameters to fit. Or we can assume that $\nu(t)$ is a general polynomial of degree $k$, so we have $4+k$ parameters to fit. Similarly, we can assume that $\nu(t)$ is some function, {\em e.g.}, $\nu(t)=\nu_0~\cos(\omega~t)$, in which case we have five parameters to fit. {\em Etc.}

{}Alternatively, we can follow a different procedure, which we set forth here. Instead of trying to fit $\nu(t)$, we can first fit the three parameters $z$, $\beta$ and $r_0$ ({\em e.g.}, via the least squares method -- see below) {\em assuming} $\nu(t)\equiv 0$, and then attribute the difference between the so-fitted model yield curve $R_m(t, T)$ and the empirical yield curve $R_e(t,T)$ to nontrivial $\nu(t)$. We will refer to this difference as the ``residual" yield: $R_r(t,T) \equiv R_e(t,T) - R_m(t,T)$. In fact, looking at (\ref{T-bond}) it is evident that for nontrivial $\nu(t)$ it is more convenient to fit $z$ (which is nothing but $r_t$), $\beta$ and $\chi(t)$ (as opposed to $r_0$). So, below, after we fit $z$, $\beta$ and $r_0$ in the $\nu(t)\equiv 0$ case, we will use the so-obtained $r_0$ as $\chi(t)$ in the nontrivial $\nu(t)$ case.\footnote{\, Equivalently, we can simply set $t=0$, so $\chi(t) = r_0$.} Then we have the following simple formula for the ``residual" yield:
\begin{equation}
 R_r(t,T) = {\eta(t,T)\over{T-t}}
\end{equation}
so we can read $\eta(t,T)$ off the empirical yield curve once we fit $R_m(t,T)$, which will give us approximate\footnote{\, Typically, there are not that many maturities available, so reconstructing $\chi(s) = \partial_s\eta(t, s) + \chi(t)$ and $\nu(s) = \dot{\chi}(s)$ would involve piecewise polynomial splines.} shapes of $\chi(s)$ and $\nu(s)$ for $t\leq s\leq T$.

\subsubsection{An Illustrative Example: Japanese Government Bonds}

{}For illustrative and comparison purposes, we have used the zero-coupon $T$-bond pricing formula (\ref{T-bond})\footnote{\, The Airy functions $Ai(y)$ and $Bi(y)$ and their zeroes are built-in within the ``gsl" package in R. Their integrals are not, so we evaluated them (see (\ref{int4}), Appendix \ref{appA}).} in the Ho and Lee model on a semi-infinite line with $\nu(t)\equiv 0$ (``Model-2") to fit the Japanese Government Bond data used in (Gorovoi and Linetsky, 2004) to calibrate Black's model of interest rates as options (Black, 1995)\footnote{\, A similar model was independently discussed by Rogers (1995).} with the underlying ``shadow rate" following the Vasicek model (``Model-1"). As in (Gorovoi and Linetsky, 2004), we fit the model by minimizing the root mean squared error (RMSE) between the empirical yield curve (Table \ref{table1}, column 4) and the Model-2 yield curve (see above).

{}Table \ref{table1} summarizes our results. The calibrated model parameters in Model-2 are\footnote{\, All dimensionful parameters are quoted in the units of 1 year. Note that $r_t$, $r_0$, $z$, $\beta$ and $E_n$ have dimension $(\mbox{time})^{-1}$, $\sigma$ has dimension $(\mbox{time})^{-3/2}$, and $W_t$ has dimension $(\mbox{time})^{1/2}$.} $z \approx -0.00184$, $\beta \approx 0.0924$ (which implies $\sigma \approx 0.0397$), $r_0 \approx -0.05834$. This implies that the initial short-rate value $r_t=z$ (here $t=02/03/2002$) is essentially 0 (within the 2-digit precision of the underlying bootstrapped yield data in column 4 of Table \ref{table1}). On the other hand, in Model-2 the lower bound on the short-rate process is $r_-=r_0\approx -5.834\%$, which is close to the initial value $-5.12\%$ of the ``shadow rate" found in (Gorovoi and Linetsky, 2004) for Model-1. However, due to the discrete spectrum in Model-2, the fact that $r_-$ is negative is not particulary informative. The short-rate modes that contribute into the bond (and other) prices are given by $\chi_n = r_0 + E_n$ (recall that in this case $\chi(t)=r_0$, so $\chi_n(t)$ are constant), which are positive in this model. The first ten values of $\chi_n$ are given in column 1, Table \ref{table3}. It is clear that at long maturities only a few lowest-lying levels have significant contributions into (\ref{T-bond}). However, at short maturities a significant number of levels must be included.\footnote{\, In our computation it was (more than) sufficient to truncate the series in (\ref{T-bond}) at $n=300$.}

{}The Model-2 and empirical yields are plotted in Figure 1. The fit in Model-2, which has fewer (to wit, 3) parameters than Model-1,\footnote{\, In Model-1 the underlying ``shadow rate" process $X_t$ follows the Vasicek model (\ref{OU}) with constant $\sigma$, $\theta$ and $\alpha$, which together with $z = r_t$ give 4 parameters. The fifth parameter is the location of the sewing point $r_-$, {\em i.e.}, $r_t = r_- + (X_t)^+$. In (Gorovoi and Linetsky, 2004) it is set to zero, $r_-=0$. However, here too the spectrum can be nonnegative for a range of $r_- < 0$.} is actually better than in Model-1. The RMSE between the empirical yield curve (Table \ref{table1}, column 4) and the Model-1 yield curve (Table \ref{table1}, column 5) $\mbox{RMSE(Model-1)}\approx 6.37\times 10^{-4}$. The RMSE between the empirical yield curve and the Model-2 yield curve (Table \ref{table1}, column 6) $\mbox{RMSE(Model-2)}\approx 5.91\times 10^{-4}$. This implies that the mean-reverting feature in Model-1 (which introduces two additional parameters) apparently does not improve the fit. We have also computed a straightforward (not piecewise spline) cubic fit with the intercept, which amounts to fitting a general cubic polynomial into the empirical yield curve using a linear model. The RMSE between the empirical yield curve and the cubic fit yield curve (Table \ref{table1}, column 7) $\mbox{RMSE(Cubic)}\approx 6.60\times 10^{-4}$. From Table \ref{table1}, column 7 it is evident that both Model-1 and Model-2 are significantly better than the cubic fit. The cubic fit works very well at long maturities, but fails badly at short maturities. This is not surprising considering that, as mentioned above, at long maturities only the lowest few modes contribute, whereas at short maturities a large number of modes do. Furthermore, this suggests that a good model fit (both for Model-1 and Model-2) may not be a universal feature and may not persist to other data, in which case the ``residual" yield might have to be attributed to nontrivial drift. The ``residual" drift in Model-2 is plotted in Figure 2.

\subsubsection{An Illustrative Example: Recent US Treasury Yield Curve}

{}Table \ref{table2} summarizes the Model-2 fit for a recent US Treasury yield curve. The calibration procedure is the same as above. Column 3 of Table \ref{table2} corresponds to the Model-2 fit based on all maturities $T$; the empirical and Model-2 yields are plotted in Figure 3, and the ``residual" yield is plotted in Figure 4; the first 10 values of the $\chi_n$ spectrum are given in Table \ref{table3}, column 2. Column 4 of Table \ref{table2} corresponds to the Model-2 fit based only on the maturities $T\geq\mbox{1 yr}$; the empirical and Model-2 yields are plotted in Figure 5, and the ``residual" yield is plotted in Figure 6; the first 10 values of the $\chi_n$ spectrum are given in Table \ref{table3}, column 3. For comparison purposes, columns 5 and 6 of Table \ref{table2} contain straightforward cubic fits into the empirical data for all maturities and $T\geq\mbox{1 yr}$ maturities only, respectively. We have the following RMSE for the above fits: $\mbox{RMSE(Model-2, All $T$)}\approx 1.99\times 10^{-3}$; $\mbox{RMSE(Model-2, $T\geq\mbox{1 yr}$)}\approx  4.91\times 10^{-4}$; $\mbox{RMSE(Cubic, All $T$)}\approx  5.07\times 10^{-4}$; $\mbox{RMSE(Cubic, $T\geq\mbox{1 yr}$)}\approx 5.39 \times 10^{-4}$. The Model-2 fit, which assumes vanishing drift, is better than the cubic fit for longer maturities, but not for shorter ones.

{}Looking at the results it is clear that Model-2 provides a very good fit for maturities $T\geq\mbox{1 yr}$, but not for short maturities $T < \mbox{1 yr}$. For these short maturities, the ``residual" yield attributed to nontrivial drift is large and cannot be neglected. Moreover, apparently, the drift is not even approximately constant -- for a constant drift $\nu(t)\equiv \nu_0$ we would have the ``residual" yield of the form $R_r(t,T) = (\nu_0/2)\left(T-t\right)$. So, the drift appears to have nontrivial time dependence.

\subsection{Bond Pricing on an Interval}

{}For the sake of completeness, let us briefly discuss zero-coupon $T$-Bond pricing in the Ho and Lee model on an interval. We have
\begin{equation}\label{c2}
 c_n = \int_0^L dx~\psi_n(x) = {\pi\over\alpha}~a_n\left[\phi_n(0)~Gi^\prime(-e_n) - \phi_n(L)~Gi^\prime(\alpha~L-e_n)\right]
\end{equation}
where we have used (\ref{int2}) and (\ref{int3}) (see Appendix \ref{appA}), and $\phi_n(x)$ is defined in (\ref{box5}).

{}The $T$-bond pricing function therefore is given by the following simple formula:
\begin{eqnarray}
 &&v(z, t, T) = e^{-\eta(t,T)}\sum_{n=1}^\infty b_n~\phi_n\left({z - \chi(t)\over \sigma}\right) e^{-\chi_n(t)\left(T-t\right)}\label{T-bond1}\\
 &&b_n \equiv \pi~{{\phi_n(0)~Gi^\prime(-e_n) - \phi_n(L)~Gi^\prime(\alpha~L-e_n)}\over {e_n~\phi_n^2(0) + \left(\alpha~L - e_n\right)\phi_n^2(L)}}
\end{eqnarray}
where $\chi_n(t) = \chi(t) + E_n = \chi(t) + \beta~e_n$, and $e_n>0$ are the roots of (\ref{box4}). Recall that $\alpha\equiv (2\sigma)^{1/3}$ and $\beta \equiv (\sigma^2/2)^{1/3}$. If we assume vanishing drift, then the Ho and Lee model on an interval has four parameters to fit: $z$, $\sigma$, $r_0$ (the lower reflecting barrier is at $r_-=r_0$) and $L$ (the upper reflecting barrier is at $r_+=r_0+L$). The calibration can be done as above with the caveat that numerical estimations involving the Airy function $Bi(y)$ are trickier as $Bi(y)$ diverges for large positive $y$ (see Appendix \ref{appA} for some useful formulas).

\section{Concluding Remarks}

{}The condition (\ref{asymp.yield}) restricts the allowed drifts $\nu(t)$. As mentioned above, the l.h.s. of this condition should be some positive number, which we will denote by $R_{min}$. Then we have\footnote{\, This holds for general $r_t = V(W_t) + \chi(t)$. If $V(x)$ is confining ({\em e.g.}, $V(x)=\omega^2~x^2/2$), then there is no need for a reflecting boundary as the spectrum is discrete and bounded from below.}
\begin{equation}
 \chi_*(t) + \chi(t) + E_1 \geq R_{min}
\end{equation}
Let us now consider constant drift $\nu(t) \equiv \nu$. We have $\chi(t) = r_0 + \nu~t$, so $\eta(t,T) = \nu\left(T-t\right)^2/2$ and $\chi_*(t)$ is $+\infty$ if $\nu>0$ and $-\infty$ if $\nu < 0$. This implies that constant negative drift is not allowed.\footnote{\, In the solution (\ref{const.drift}) corresponding to introducing a reflecting barrier for $r_t$ (as opposed to $W_t$) constant negative drift {\em a priori} is allowed as $r_t$ is reflected when it hits its lower boundary $r_*$.} This is evident from the fact that when the drift is constant we have $r_t = V(W_t) + \nu~t$. In fact, positive constant $\nu$ is not realistic either because it would drive $r_t$ to large values. A reasonable assumption then is that $\chi_*(t)$ should be finite, which implies that
\begin{equation}
 \lim_{s\rightarrow\infty}\left|\chi(s)\right| \leq {\widetilde \chi}
\end{equation}
where ${\widetilde\chi}$ is finite.

{}Another point worth commenting on is that, while for time-homogeneous cases ({\em i.e.}, for time-independent parameters) we can consider ``option-like" models with $r_t = (X_t)^+$ and the underlying ``shadow" rate $X_t$ following processes such as Vasicek, CIR, {\em etc.}, which have discrete spectra, $X_t$ cannot follow the Ho and Lee model as the latter has continuous spectrum.\footnote{\, Here one can wonder if it would make sense to consider ``option-like" models for $W_t$. However, if we replace $W_t$ by $(W_t)^+$ in $r_t = V(W_t) + \chi(t)$, the spectrum will be continuous.} In contrast, we can have a reflecting barrier in the Ho and Lee model. Such a barrier can be introduced for $W_t$ for a time-dependent drift (subject to the above conditions), and also on $r_t$ for constant drift, which is negative in the case with only the lower barrier, but can be positive in the case where both lower and upper barriers are present. However, constant drift is limited in its applicability as can be seen from the U.S. Treasury data we discussed above.

\appendix

\section{Some Properties of Airy and Scorer Functions}\label{appA}

{}In this appendix we collect some properties of the Airy and Scorer functions used in the main text. Unless stated otherwise, these properties are taken from (Abramowitz and Stegun, 1964).

{}$\bullet$ Integral identity:
\begin{equation}\label{int1}
 \int dy~f_1(y)~f_2(y) = y~f_1(y)~f_2(y) - f_1^\prime(y)~f_2^\prime(y)
\end{equation}
where each $f_1$ and $f_2$ can be either $Ai(y)$ or $Bi(y)$. Eq. (\ref{int1}) can be obtained via integration by parts and using the Airy equation $f_i^{\prime\prime}(y) = y~f_i(y)$, $i=1,2$.

{}$\bullet$ Integral identities:
\begin{eqnarray}\label{int2}
 &&{1\over\pi} \int_y^\infty dw~Ai(w) = Ai(y)~Gi^\prime(y) - Ai^\prime(y)~Gi(y)\\
 &&{1\over\pi} \int_0^y dw~Bi(w) = Bi^\prime(y)~Gi(y) - Bi(y)~Gi^\prime(y)\label{int3}\\
 &&\int_0^\infty dw~Ai(w) = {1\over 3}\label{int5}
\end{eqnarray}
where $Gi(y)$ is a Scorer function. Let
\begin{eqnarray}
 &&{\cal A}(-y)\equiv \int_{-y}^0 dw~Ai(w)\\
 &&{\cal B}(-y)\equiv \int_{-y}^0 dw~Bi(w)
\end{eqnarray}
where $y > 0$. The integrals ${\cal A}(-y)$ and ${\cal B}(-y)$ can be evaluated numerically. It follows from  (\ref{int5}) that
\begin{equation}\label{int4}
 \int_{-y}^\infty dw~A(w) = {1\over 3} + {\cal A}(-y)
\end{equation}
where $y > 0$. It further follows from (\ref{int2}) and (\ref{int3}) that
\begin{equation}
 \pi~Gi^\prime(-y) = {{\left[{1\over3}+ {\cal A}(-y)\right] Bi^\prime(-y) - {\cal B}(-y)~Ai^\prime(-y)}\over{Ai(-y)~Bi^\prime(-y) - Bi(-y)~Ai^\prime(-y)}}
\end{equation}
which allows to evaluate $Gi^\prime(-y)$ numerically for $y > 0$. In some cases, the integral representation for $Gi(y > 0)$ given in (Gil, Segura and Temme, 2001) may be useful for evaluating $Gi^\prime(y > 0)$.

{}$\bullet$ Leading asymptotic behavior
\begin{eqnarray}
 &&Ai(-y) \sim {y^{-1/4}\over\sqrt{\pi}}~\sin\left({2\over 3}~y^{3/2} +{\pi\over 4}\right)\\
 &&Gi^\prime(-y) \sim {y^{1/4}\over\sqrt{\pi}}~\sin\left({2\over 3}~y^{3/2} +{\pi\over 4}\right)
\end{eqnarray}
at $y\gg 1$.

{}$\bullet$ Leading asymptotic behavior at large $n$:
\begin{eqnarray}
 &&\xi_n \sim -\left({3\pi\over 8}~[4n-3]\right)^{2/3}\\
 &&Ai\left(\xi_n\right)\sim {(-1)^{n+1}\over \sqrt{\pi}} \left|\xi_n\right|^{-1/6}\\
 &&Gi^\prime\left(\xi_n\right)\sim {(-1)^{n+1}\over \sqrt{\pi}} \left|\xi_n\right|^{1/6}
\end{eqnarray}
where $\xi_n$ ($0 > \xi_1 > \xi_2 > \dots$) are the zeros of $Ai^\prime(y)$.

\section{Reflecting Barriers for the Short-rate Process}\label{appB}

{}In this section we discuss the issues associated with introducing reflecting boundaries directly for the short-rate process $r_t$ as opposed to the underlying Brownian motion process $W_t$. Our starting point is (\ref{pricing.PDE}) together with the terminal condition $v(z,T,T)={\widetilde Y}(z)$. When $\sigma(r_t,t)$ and $\nu(r_t,t)$ have no explicit time dependence, reflecting barriers for $r_t$ can be introduced as in, {\em e.g.}, (Goldstein and Keirstead, 1997) and (Gorovoi and Linetsky, 2004).\footnote{\, In this case, (\ref{pricing.PDE}) can be transformed into the Schr\"odinger equation with time-independent potential (a.k.a. Liouville normal form) via a Liouville transformation.} Even if $\sigma(r_t,t)$ has no explicit time dependence, for general, explicitly time-dependent $\nu(r_t,t)$ things are trickier. If fact, for our purposes here, to illustrate our main point, it will suffice to consider the case where $\sigma(r_t,t)\equiv\sigma$ is constant and $\nu(r_t,t)\equiv \nu(t)$ is independent of $r_t$, but is a deterministic function of $t$. Eq. (\ref{pricing.PDE}) then reduces to
\begin{equation}\label{PDE.v}
 \nu(t)~ \partial_z v(z,t,T)+ \partial_t
 v(z,t,T)+{1\over 2}\sigma^2~\partial^2_z v(z,t,T)-z~v(z,t, T)=0
\end{equation}
An analogue of the Liouville transformation in this case is
\begin{equation}
 v(z, t, T)\equiv\exp\left(-{\nu(t)~z\over\sigma^2}\right)~u(z, t, T)
\end{equation}
Then the PDE for $u(z, t, T)$ reads:
\begin{equation}\label{PDE.u}
 \partial_t
 u(z,t,T)+{1\over 2}\sigma^2~\partial^2_z u(z,t,T)-\left[z\left(1+{\dot{\nu}(t)\over\sigma^2}\right) + {\nu^2(t)\over 2\sigma^2}\right]u(z,t, T)=0
\end{equation}
This is the Schr\"odinger equation with time-dependent potential -- unless $\nu(t)$ is constant, that is. When $\nu(t)$ is a linear function of $t$ so that $\dot{\nu}(t)\equiv \rho$ is constant, the potential is time-dependent but factorized, so in this special case the solution to (\ref{PDE.u}) is given by
\begin{equation}
 u(z, t, T) =\exp\left(\int_0^t ds~{\nu^2(s)\over 2\sigma^2}\right) w(z,t,T)
\end{equation}
where $w(z,t,T)$ solves the following Schr\"odinger equation with a linear potential:
\begin{equation}\label{PDE.w}
 \partial_t
 w(z,t,T)+{1\over 2}\sigma^2~\partial^2_z w(z,t,T)-z\left(1+{\rho\over\sigma^2}\right) w(z,t, T)=0
\end{equation}
which can be solved as in the main text; provided, however, that the boundary condition can be consistently specified.

{}This is where a difficulty arises irrespective of whether $\nu(t)$ is a linear function of $t$ or not -- so long as it is not constant, that is. Thus, if we impose a reflecting (Neumann) boundary condition for some $r_t=r_*$
\begin{equation}
 \left.\partial_z v(z,t,T)\right|_{z=r_*}=0
\end{equation}
then for $u(z, t, T)$ we have the following Robin boundary condition:
\begin{equation}\label{Robin}
  \left.\partial_z u(z,t,T)\right|_{z=r_*}=\left.{\nu(t)\over\sigma^2}~u(z,t,T)\right|_{z=r_*}
\end{equation}
For constant $\nu(t)$ we can perform the separation of variables. However, even for linear $\nu(t)$ the straightforward separation of variables procedure cannot be applied and the problem becomes more difficult to solve.

{}Let us briefly outline the solution for constant $\nu(t)\equiv \nu$, primarily to illustrate the difference with the corresponding solution in the case of a reflecting boundary for the underlying Brownian motion we discuss in the main text. We have\footnote{\, Here we focus on the solution with only one (lower) boundary. The solution with both (lower and upper) boundaries can also be readily constructed and involves both $Ai(y)$ and $Bi(y)$.}
\begin{eqnarray}\label{const.drift}
 &&v(z, t, T) = \psi(z/\beta, T-t)\\
 &&\psi(y,\tau) = e^{-\gamma y} \sum_{n=1}^\infty d_n~ Ai\left(y - e_n\right)~e^{-\lambda_n\tau}
\end{eqnarray}
where $\lambda_n \equiv \beta\left(\gamma^2 + e_n\right)$, $\beta\equiv (\sigma^2/2)^{1/3}$, $\gamma\equiv \nu/(2\sigma^4)^{1/3}$, and $e_n$ are the roots of the equation
\begin{equation}
 Ai^\prime\left(y_* - e_n\right) = \gamma~Ai\left(y_* - e_n\right)
\end{equation}
which is a consequence of the boundary condition (\ref{Robin}) corresponding to a reflecting boundary\footnote{\, The reflecting boundary in (Goldstein and Keirstead, 1997) was set at $r_* = 0$.} at $r_t = r_*$, and $y_* \equiv r_*/\beta$. The coefficients $d_n$ are fixed via the terminal condition $v(z, T, T) = {\widetilde Y}(z)$. For a zero-coupon $T$-bond we have ${\widetilde Y}(z) \equiv 1$ and
\begin{eqnarray}
 &&d_n = {{\int_{y_*}^\infty dy~e^{\gamma y} Ai(y - e_n)}\over{\int_{y_*}^\infty dy \left[Ai(y - e_n)\right]^2}} = e^{\gamma e_n}~{{I(\gamma, e_n - y_*) + {\widetilde I}(\gamma)}\over{\left(e_n - y_* - \gamma^2\right)\left[Ai(y_* - e_n)\right]^2}}\\
 &&I(\gamma, y) \equiv \int_{-y}^0 dw~e^{\gamma w} Ai(w)\\
 &&{\widetilde I}(\gamma) \equiv \int_0^\infty dw~e^{\gamma w} Ai(w) = \nonumber\\
 &&\,\,\,\,\,\,\,e^{\gamma^3/3}\left[{1\over 3} + {_1F_1\left({1\over3};{4\over3};-{\gamma^3\over 3}\right) \gamma \over 3^{4/3}~\Gamma(4/3)} - {_1F_1\left({2\over3};{5\over3};-{\gamma^3\over 3}\right) \gamma^2 \over 3^{5/3}~\Gamma(5/3)}\right]\label{hyper}
\end{eqnarray}
where $\Gamma(y)$ is the Gamma function, and $_1F_1(a;b;y)$ is a generalized hypergeometric function. In (\ref{hyper}) we have used a Laplace transform of the Airy function $Ai(y)$, see Eq. (9.10.14) in (DLMF, 2015). In practice, if $|\gamma|$ is not large, then the integral ${\widetilde I}(\gamma)$ can be evaluated by truncating the series
\begin{equation}
 {\widetilde I}(\gamma) = {1\over 3} \sum_{n=0}^\infty {\left(\gamma/3^{1/3}\right)^n\over \Gamma(n/3+1)}
\end{equation}
which follows from the Mellin transform of $Ai(y)$, see Eq. (9.10.17) in (DLMF, 2015). Asymptotically, this series behaves as the Taylor expansion of $\left(1/3\right)\exp(\gamma^3/3)$.

\subsection*{Acknowledgments}
{}I would like to thank Eyal Neuman for discussions on Brownian motion with reflecting barriers, which prompted me to think about this topic.

\newpage
\begin{table}[ht]
\noindent
\caption{The calibration results for the zero-coupon $T$-bond pricing formula (\ref{T-bond}) in the Ho and Lee model on a semi-infinite line with vanishing drift (``Model-2"). The first five columns are taken from (Gorovoi and Linetsky, 2004). The first three columns give the Bloomberg data as of $t=02/03/2002$ for the Japanese Government Bonds: coupon, maturity $T$ and price. The fourth column gives the corresponding bootstrapped zero-coupon bond yields. The fifth column gives the calibrated model yields calculated in (Gorovoi and Linetsky, 2004) for Black's model of interest rates as options (Black, 1995) with the underlying ``shadow rate" following the Vasicek model (``Model-1"). The sixth column gives calibrated model yields we calculated based (\ref{T-bond}). The calibrated model parameters are $z \approx -0.00184$, $\beta \approx 0.0924$ (which implies $\sigma \approx 0.0397$), $r_0 \approx -0.05834$. For illustrative purposes we have kept 3 digits in the sixth column. In the seventh column we have included the yield curve from fitting a general cubic polynomial into the empirical data in the fourth column.} 
\begin{tabular}{l l l l l l l} 
\\
\hline\hline 
Coupon & Maturity & Price & Empirical & Model-1 & Model-2 & Cubic Fit\\
& & & 0-Coupon & Yield (\%) & Yield (\%) & Yield (\%)\\
& & & Yield (\%) &   &   &  \\[0.5ex] 
\hline 
4.2 &  3/20/2003 & 104.648 & 0.02 & 0.03 & 0.023 & -0.106\\
3.4 &  3/22/2004 & 106.900 & 0.14 & 0.17 & 0.106 & 0.156\\
4.4 &  3/21/2005 & 112.729 & 0.30 & 0.36 & 0.338 & 0.395\\
3.1 &  3/20/2006 & 110.481 & 0.54 & 0.57 & 0.571 & 0.615\\
2.6 &  3/20/2007 & 109.326 & 0.76 & 0.78 & 0.788 & 0.818\\
1.9 &  3/20/2008 & 105.578 & 0.98 & 0.98 & 0.988 & 1.005\\
1.9 &  3/20/2009 & 104.723 & 1.24 & 1.16 & 1.169 & 1.175\\
1.7 &  3/22/2010 & 102.521 & 1.40 & 1.33 & 1.333 & 1.331\\
1.4 &  3/21/2011 & 99.314  & 1.51 & 1.48 & 1.481 & 1.472\\
1.5 & 12/20/2011 & 99.997  & 1.53 & 1.59 & 1.584 & 1.569\\
3.8 &  9/20/2016 & 123.287 & 2.11 & 2.09 & 2.084 & 2.040\\
2.1 & 12/20/2021 & 98.411  & 2.29 & 2.44 & 2.434 & 2.360\\
2.4 & 11/20/2031 & 94.810  & 2.88 & 2.79 & 2.801 & 2.869\\ [1ex] 
\hline 
\end{tabular}
\label{table1} 
\end{table}

\begin{table}[ht]
\noindent
\caption{The calibration results for the zero-coupon $T$-bond pricing formula (\ref{T-bond}) in the Ho and Lee model on a semi-infinite line with vanishing drift (``Model-2"). The first two columns are taken from the U.S. Treasury website (U.S. Treasury, 2015) as of $t=01/29/2015$: maturity $T$ and empirical yield $R_e(t, T)$. The third column is the Model-2 fit including all maturities, for which the calibrated model parameters are $z \approx -0.0027$, $\beta \approx 0.2516$ (which implies $\sigma \approx 0.1785$), $r_0 \approx -0.23163$. The fourth column is the Model-2 fit including only the maturities of 1 year and longer, for which the calibrated model parameters are $z \approx 0.0012$, $\beta \approx 0.2085$ (which implies $\sigma \approx 0.1346$), $r_0 \approx  -0.1879$. For illustrative purposes we have kept 3 digits in the third and fourth columns. In the fifth (all maturities) and sixths (only the maturities of 1 year and longer) columns we have included the yield curve from fitting a general cubic polynomial into the empirical data in the second column.} 
\begin{tabular}{l l l l l l} 
\\
\hline\hline 
Maturity & Empirical & Model-2 & Model-2 & Cubic Fit & Cubic Fit\\
$T$ & Yield (\%) & Yield (\%) & Yield (\%) & Yield (\%) & Yield (\%)\\
&  & (All $T$)  & ($T\geq\mbox{1 yr}$)  &  (All $T$) & ($T\geq\mbox{1 yr}$)\\[0.5ex] 
\hline 
1 mo    & 0.01 & 0.046  & ---   & -0.040 & ---\\
3 mo    & 0.03 & 0.524  & ---   & 0.016  & ---\\
6 mo    & 0.07 & -0.005 & ---   & 0.097  & ---\\
1 yr    & 0.17 & -0.151 & 0.209 & 0.252  & 0.224\\
2 yrs   & 0.51 & 0.292  & 0.464 & 0.539  & 0.521\\
3 yrs   & 0.84 & 0.733  & 0.797 & 0.795  & 0.785\\
5 yrs   & 1.28 & 1.314  & 1.295 & 1.221  & 1.224\\
7 yrs   & 1.59 & 1.629  & 1.594 & 1.544  & 1.554\\
10 yrs  & 1.77 & 1.880  & 1.845 & 1.865  & 1.878\\
20 yrs  & 2.11 & 2.175  & 2.147 & 2.086  & 2.078\\
30 yrs  & 2.33 & 2.273  & 2.249 & 2.335  & 2.336\\ [1ex] 
\hline 
\end{tabular}
\label{table2} 
\end{table}

\begin{table}[ht]
\noindent
\caption{The first 10 values $\chi_n$ contributing to the zero-coupon $T$-bond pricing formula (\ref{T-bond}) in the Ho and Lee model on a semi-infinite line with vanishing drift (``Model-2"). The first column corresponds to the Japanese Government Bond data in Table \ref{table1} (JGB). The second column corresponds to the US Treasury yield data for the fit including all maturities in column 3 of Table \ref{table2} (UST1). The third column corresponds to the US Treasury yield data for the fit including only the maturities of 1 year and longer in column 4 of Table \ref{table2} (UST2). Note that $\chi_n$ are independent of $t$ when the drift vanishes.} 
\begin{tabular}{l l l l} 
\\
\hline\hline 
$\chi_n$ & JGB (\%) & UST1 (\%) & UST2 (\%)\\[0.5ex] 
\hline 
$\chi_1$ & 3.578 & 2.470 & 2.451\\
$\chi_2$ & 24.187 & 58.562 & 48.934\\
$\chi_3$ & 38.718 & 98.111 & 81.708\\
$\chi_4$ & 51.134 & 131.906 & 109.714\\
$\chi_5$ & 62.309 & 162.321 & 134.919\\
$\chi_6$ & 72.628 & 190.407 & 158.194\\
$\chi_7$ & 82.306 & 216.749 & 180.023\\
$\chi_8$ & 91.478 & 241.713 & 200.711\\
$\chi_9$ & 100.236 & 265.549 & 220.464\\
$\chi_{10}$ & 108.646 & 288.438 & 239.432\\ [1ex] 
\hline 
\end{tabular}
\label{table3} 
\end{table}

\newpage
\begin{figure}[ht]
\centerline{\epsfxsize 4.truein \epsfysize 4.truein\epsfbox{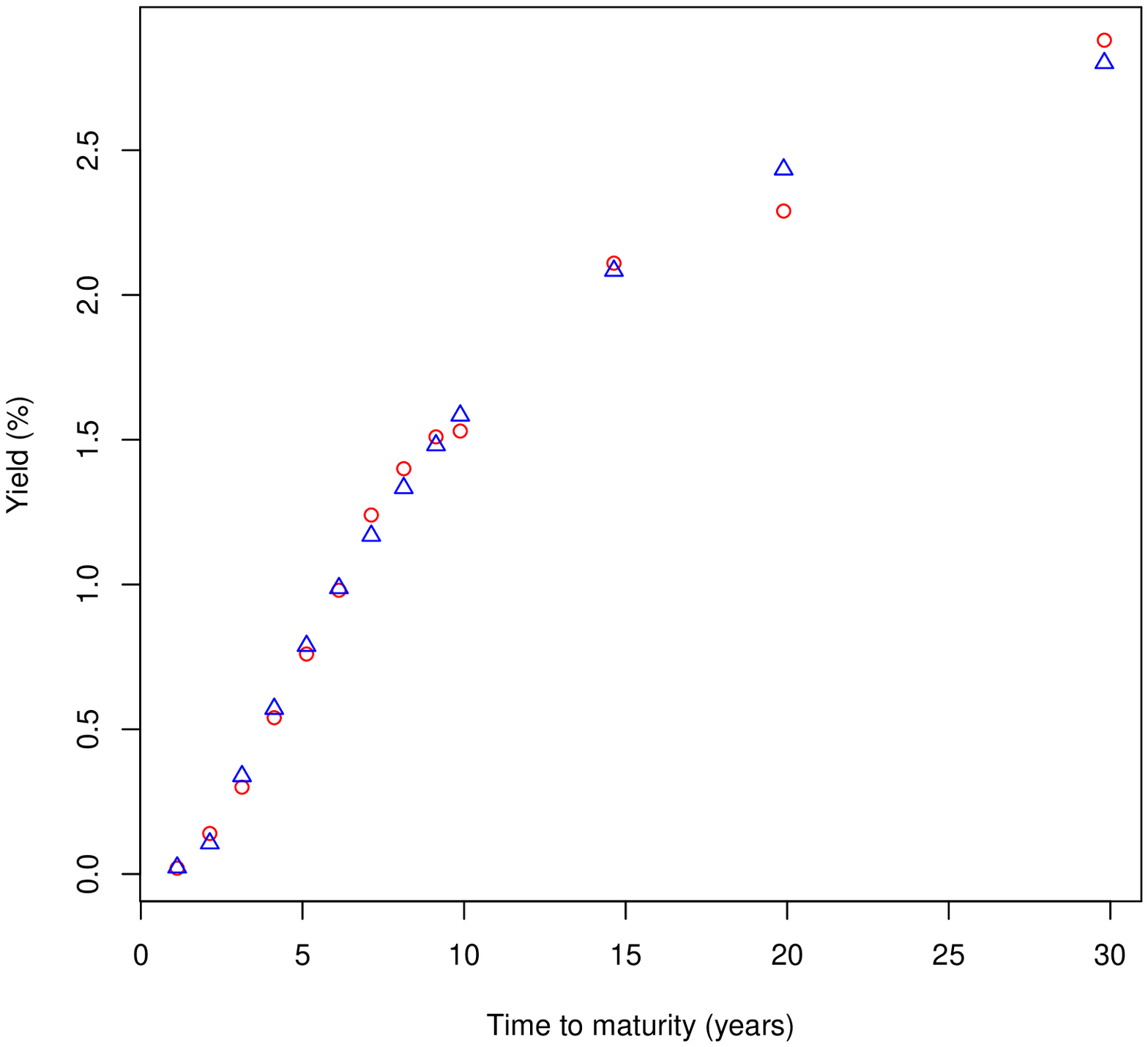}}
\noindent{\small {Figure 1. Empirical (circles) and Model-2 (triangles) yield curves corresponding to the data in Table \ref{table1}.}}
\end{figure}

\newpage
\begin{figure}[ht]
\centerline{\epsfxsize 4.truein \epsfysize 4.truein\epsfbox{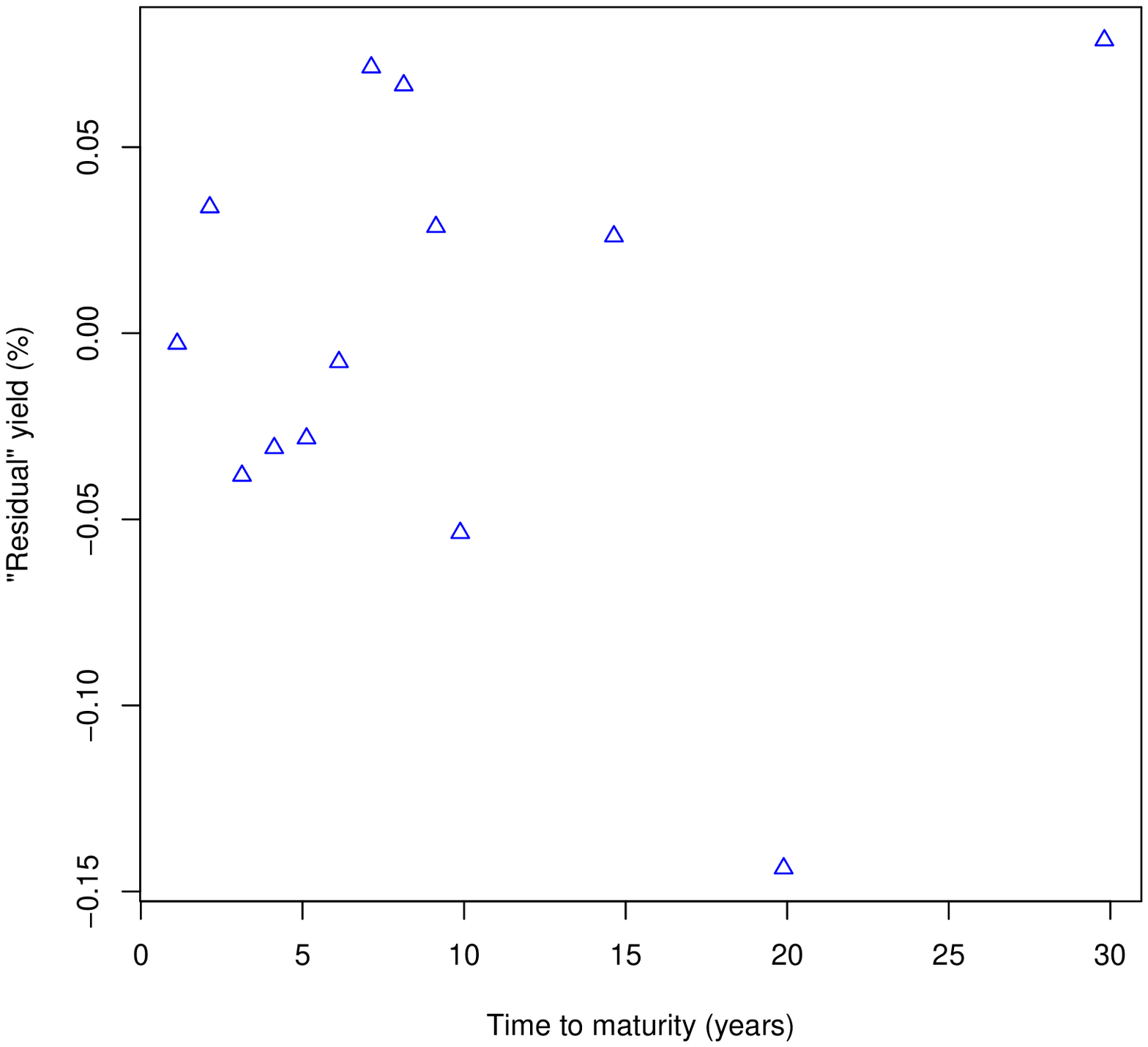}}
\noindent{\small {Figure 2. The ``residual" yield attributed to nontrivial drift in Model-2 corresponding to the data in Table \ref{table1}.}}
\end{figure}

\newpage
\begin{figure}[ht]
\centerline{\epsfxsize 4.truein \epsfysize 4.truein\epsfbox{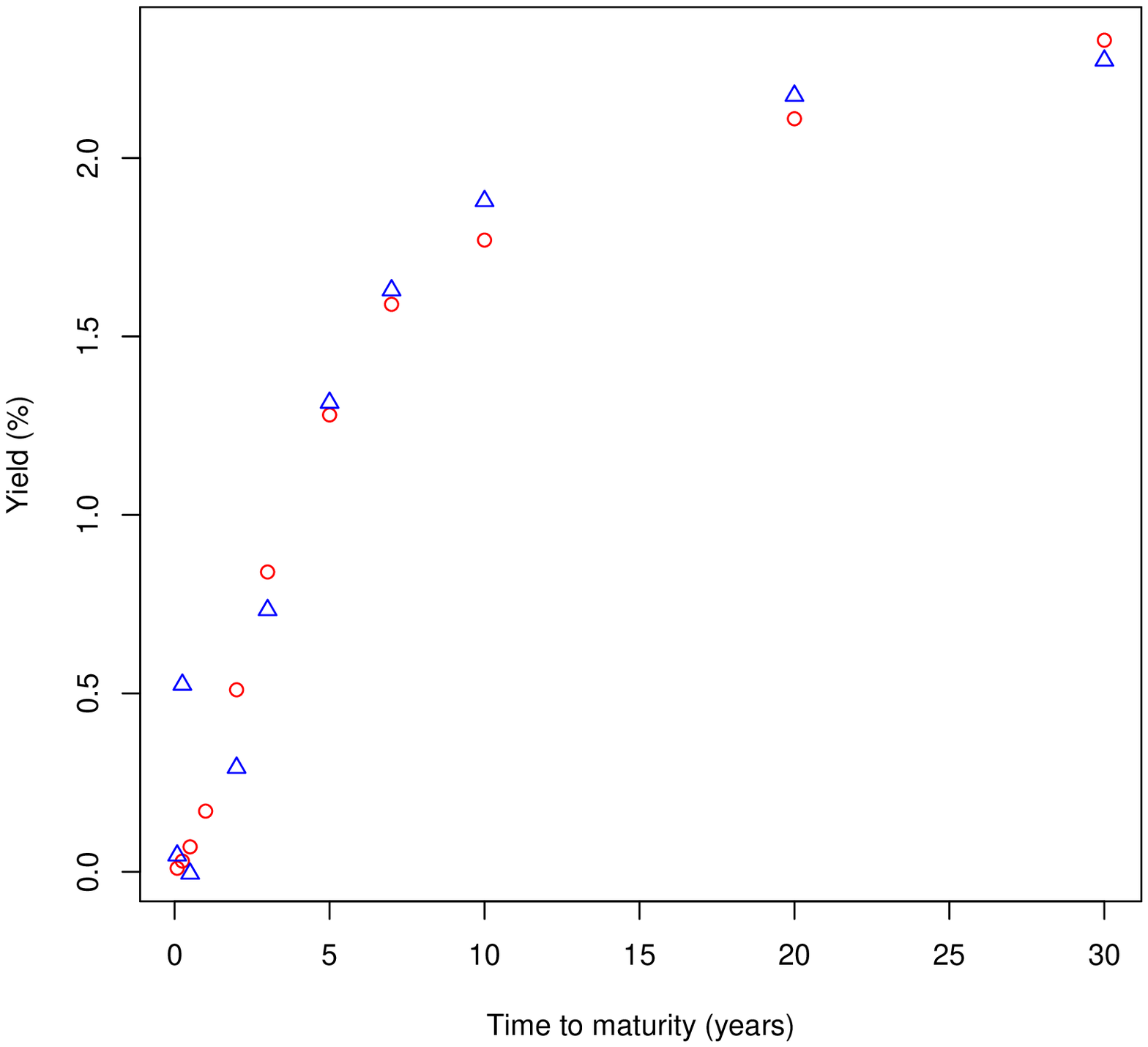}}
\noindent{\small {Figure 3. Empirical (circles) and Model-2 (triangles) yield curves corresponding to the data in Table \ref{table2} for all maturities.}}
\end{figure}

\newpage
\begin{figure}[ht]
\centerline{\epsfxsize 4.truein \epsfysize 4.truein\epsfbox{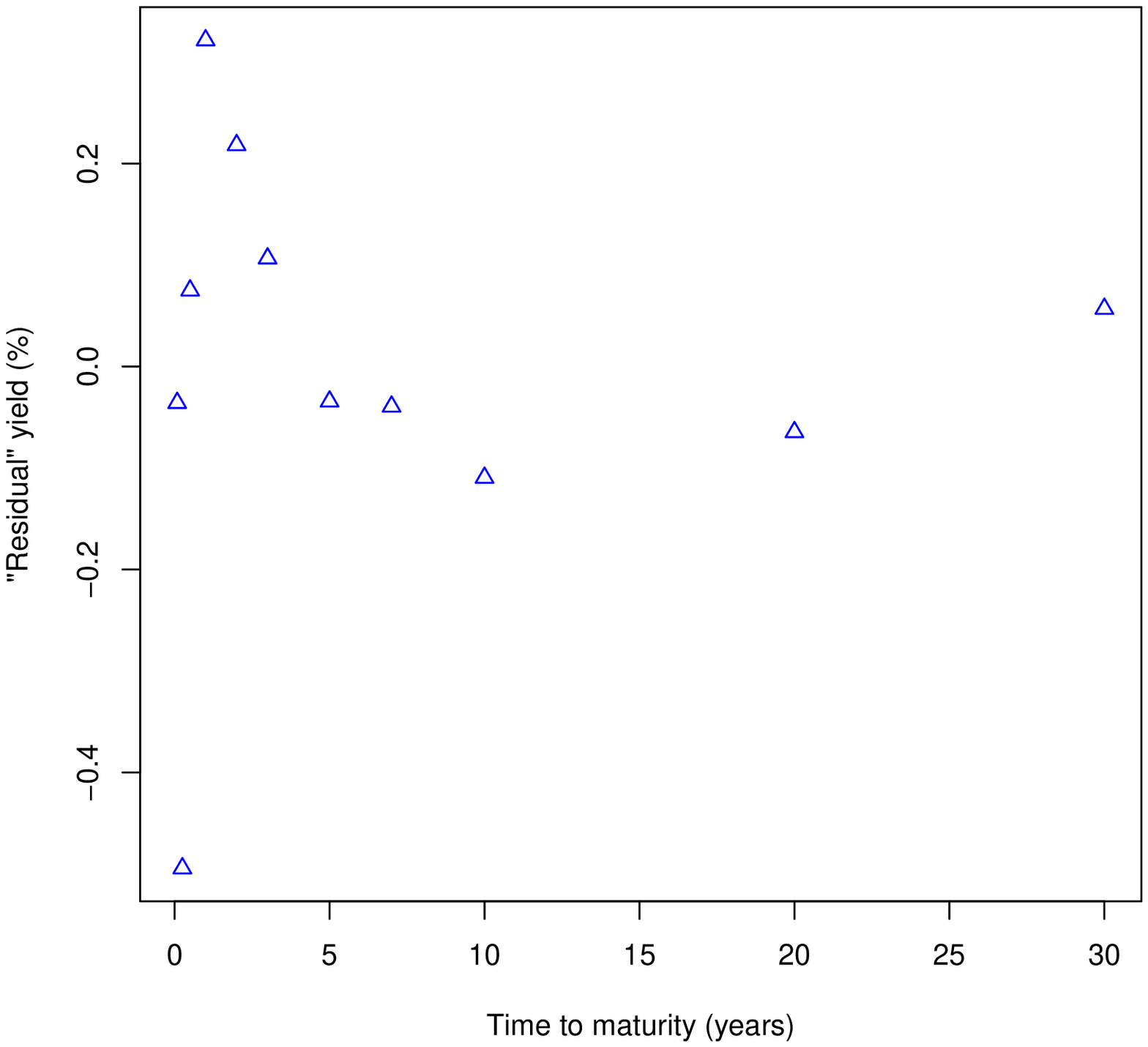}}
\noindent{\small {Figure 4. The ``residual" yield attributed to nontrivial drift in Model-2 corresponding to the data in Table \ref{table2} for all maturities.}}
\end{figure}

\newpage
\begin{figure}[ht]
\centerline{\epsfxsize 4.truein \epsfysize 4.truein\epsfbox{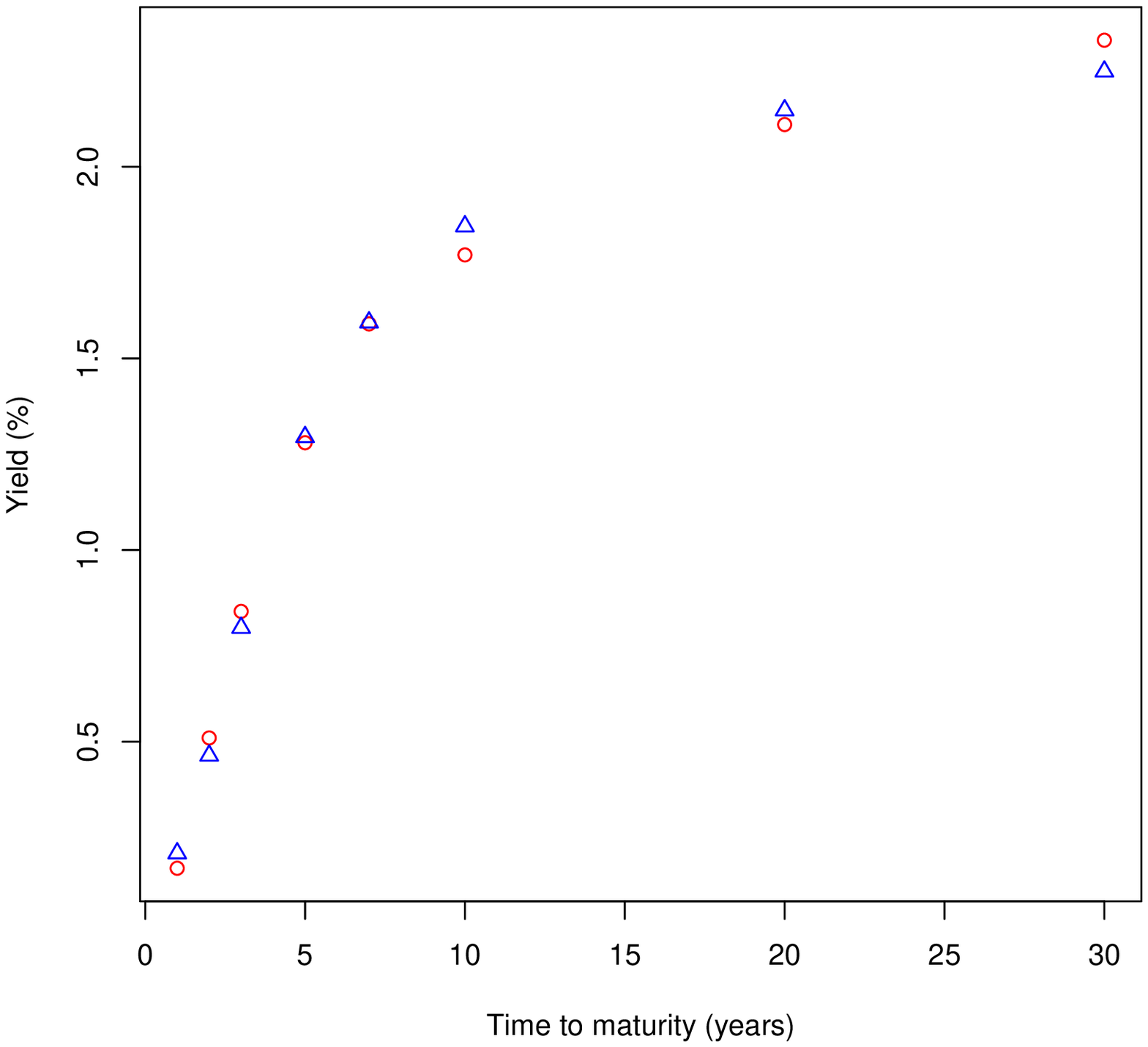}}
\noindent{\small {Figure 5. Empirical (circles) and Model-2 (triangles) yield curves corresponding to the data in Table \ref{table2} for the maturities of 1 year and longer.}}
\end{figure}

\newpage
\begin{figure}[ht]
\centerline{\epsfxsize 4.truein \epsfysize 4.truein\epsfbox{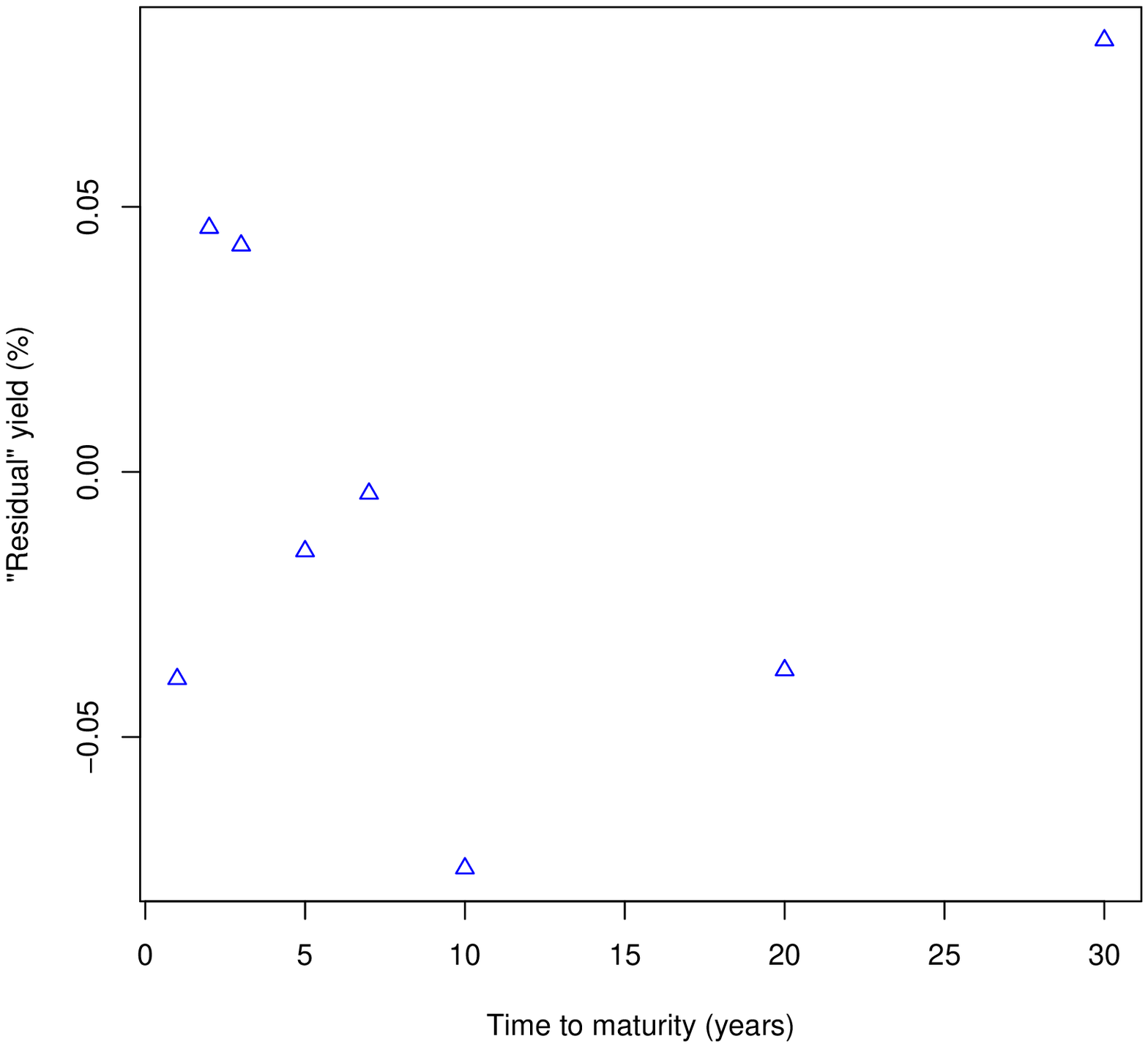}}
\noindent{\small {Figure 6. The ``residual" yield attributed to nontrivial drift in Model-2 corresponding to data in Table \ref{table2} for the maturities of 1 year and longer.}}
\end{figure}

\end{document}